\documentclass[twocolumn,preprintnumbers,amsmath,amssymb,floatfix]{revtex4}

\usepackage{graphicx}
\usepackage{dcolumn}
\usepackage{bm}
\usepackage{epsf}
\usepackage{amsmath}
\usepackage{amsfonts}

\newcommand{\DDir}{\relax{D\kern-.7em{/}}}

\newcommand{\haf}{\frac{1}{2}}
\newcommand{\inv}[1]{\frac{1}{#1}}

\newcommand{\soo}{\Rightarrow}

\newcommand{\X}{\times}

\newcommand{\bj}{\textbf{j}}

\newcommand{\be}{\begin{equation}}
\newcommand{\ee}{\end{equation}}
\newcommand{\bea}{\begin{equation*}}
\newcommand{\eea}{\end{equation*}}

\newcommand{\abs}[1]{\left\vert#1\right\vert}

\newcommand{\ave}[1]{\left\langle #1\right\rangle}

\newcommand{\nin}{\relax{\in\kern-.8em{/}}}

\newcommand{\te}{\theta}

\newcommand{\Om}{\Omega}
\newcommand{\om}{\omega}

\newcommand{\ep}{\epsilon}

\newcommand{\yr}{\mbox{ yr}}

\newcommand{\bz}{\textbf{z}}

\newcommand{\hz}{\hat\bz}

\newcommand{\bee}{\textbf{e}}
\newcommand{\bxx}{\textbf{x}}
\newcommand{\hx}{\hat \bxx}

\newcommand{\Oct}{\text{Oct}}
\newcommand{\Quad}{\text{Quad}}
\newcommand{\emt}{e_{\min}^2}

\newcommand{\per}{\text{per}}
\newcommand{\out}{\per}
\newcommand{\tsec}{t_{\rm sec}}
\newcommand{\AU}{\rm AU}

\begin{document}

\title{Long-Term Cycling of Kozai-Lidov Cycles:
Extreme Eccentricities and Inclinations Excited by a Distant Eccentric Perturber}
\author{Boaz Katz\footnote{John Bahcall Fellow, Einstein Fellow}}
\author{Subo Dong\footnote{Sagan Fellow}}
\affiliation{Institute for Advanced Study, Princeton, NJ 08540, USA}
\author{Renu Malhotra}
\affiliation{Lunar and Planetary Laboratory, The University of Arizona, Tucson, AZ 85721, USA.}
\begin{abstract}
Kozai-Lidov oscillations of Jupiter-mass planets, excited by comparable planetary or brown dwarf mass perturbers were recently shown in numerical experiments to be slowly modulated and to exhibit striking features, including extremely high eccentricities and the generation of retrograde orbits with respect to the perturber.  Here we solve this problem analytically for the case of a test particle orbiting a host star and perturbed by a distant companion whose orbit is eccentric and highly inclined.  We give analytic expressions for the conditions that produce retrograde orbits and high eccentricities.  This mechanism likely operates in various systems thought to involve Kozai-Lidov oscillations such as tight binaries, mergers of compact objects, irregular moons of planets and many others. In particular, it could be responsible for exciting eccentricities and inclinations of exo-planetary orbits and be important for understanding the spin-orbit (mis)alignment of hot Jupiters. 
\end{abstract}

\maketitle

A Keplerian orbit weakly perturbed by a distant orbiting mass may exhibit long term, large amplitude cycles in which the eccentricity and inclination change periodically \cite{Lidov62,Kozai62}. These so-called Kozai-Lidov cycles are owed to the orbit-averaged quadrupole potential of the perturber. The high eccentricity excited from initially nearly circular orbits by this mechanism is suggested to play an important role in the formation and evolution of  many astrophysical systems \cite[e.g.][]{Kozai62,Heisler86,Blaes02,Fabrycky07,Perets09,Thompson10,Naoz11}.

Kozai-Lidov oscillations of Jupiter-mass planets, excited by comparable planetary or brown dwarf mass perturbers were recently shown in numerical experiments to be slowly modulated and to exhibit striking features, including extremely high eccentricities and the generation of retrograde orbits with respect to the perturber \cite{Naoz11} (see also \cite{Ford00}).  
Those authors attributed the slow modulation of the Kozai-Lidov cycles to the relative proximity (including the octupole potential of the perturber) and comparable mass of the perturber to the perturbed planet, and the high eccentricities to chaotic evolution. These effects were argued to play a major role in affecting the properties of hot Jupiters (extra-solar, Jupiter mass planets with very short periods $\lesssim$ 10 days) and in particular in explaining the large fraction of hot Jupiters recently found to have a retrograde orbit with respect to their host's spin.

In this letter we show that the Kozai-Lidov cycles are slowly modulated non-chaotically and quite simply by the octupole potential of the perturber, which is non-vanishing when the perturber's orbit is eccentric.  This slow modulation can excite extremely high eccentricity and inclination of an initially nearly circular Keplerian orbit.  A particular consequence is that the orbit plane can flip to retrograde with respect to the total angular momentum of the system \cite{Naoz11}.  These effects occur in the test particle approximation in which the mass of the perturber is much bigger than that of the planet, and will thus be important for stellar binary perturbers of planets. We describe this long term evolution of the Kozai-Lidov cycles by deriving and analytically solving the effective equations, averaged over the Kozai cycles.

\paragraph*{Secular Equations}
Consider a test particle on a Keplerian orbit (semi-major axis $a$ and
 eccentricity $e$) subject to perturbation by a distant mass $M_{\per}$ on an orbit ($a_\per$, $e_\per$) around the same central mass $M$. The coordinate system is defined using the perturber's orbit, with the z-axis chosen to be in the direction of the angular momentum vector and the x-axis pointing to the pericenter. It is useful to parametrize the test particle's orbit by two dimensionless vectors: 
 $\bj=\mathbf{J}/\sqrt{GMa}$, where $\mathbf{J}$ is the specific angular 
 momentum vector and $G$ is the universal constant of gravitation;  $\bee$, a vector pointing in the direction of the pericenter with magnitude $e$. The orientation of $\bj$ is defined by the inclination $i$ with respect to $\hz$ and by the longitude of ascending node $\Om$ (angle between $\hz\times \bj$ and $\hx$), as follows
${
\bj=j(\sin i\sin\Om,~-\sin i\cos\Om,~\cos i).
}$
Usually, the orientation of $\bee$ is set by additionally specifying the argument of pericenter $\om$ (angle between $\bee$ and $\hz\times \bj$). Here we define the orientation of $\bee$ by the co- latitudinal angle $0\leq i_e\leq\pi$ (angle between $\hz$ and $\bee$), and longitude $\Om_e$ (angle between the projection of $\bee$ on the $xy$ plane and $\hx$) by 
\begin{equation}\label{eq:ei_eOm_e}
\bee=e(\sin i_e\cos\Om_e,~ \sin i_e\sin\Om_e,~\cos i_e).
\end{equation}
It turns out that for the cases considered, $\Om_e$ is slowly varying and is useful for describing the long term behavior of the system. 
 
The secular orbital evolution of the test particle is determined by 
double time-averaging the perturbing potential  $\Phi_{\per}$ over the 
orbital periods of the test particle and the perturber.  The averaged potential 
expanded to the octupole order (3rd order in $a/a_\per$) is given by 
$
\ave{\Phi_{\per}} = \Phi_0 \phi = \Phi_0 (\phi_{\Quad}+\ep_{\Oct}\phi_{\Oct}) 
$,
 where the dimensionless averaged potential $\phi$ is expressed as the sum of two components (quadrupole and octupole), 
\begin{eqnarray}\label{eq:phi_Q_phi_oct}
\phi_{\Quad} &=&\frac{3}{4}( \haf j_z^2 +e^2-\frac52e_z^2 -\frac16)\label{eq:phi_Koz},\\
\phi_{\Oct}&=&\frac{75}{64}\left[e_x(\inv{5}-\frac85e^2+7e_z^2-j_z^2)-2e_zj_xj_z\right],\\
\end{eqnarray}
and the normalization parameters are
\begin{equation}
\Phi_0 = \frac{GM_{\out}a^2}{{a_{\out}^3{(1-e_{\out}^2})}^{3/2}}, ~~~~\ep_{\Oct}=\frac{a}{a_{\out}}\frac{e_{\out}}{1-e_\out^2}.
\end{equation}

In the secular approximation, $a$ and $\phi$ are constant with time while 
$\bj$ and $\bee$ evolve according to the following equations of motion \cite{Milankovich39,Allan63,Tremaine09},
\begin{equation}\label{eq:NLEOM}
\frac{d\bj}{d\tau}=\bj\X\nabla_{\bj}\phi+\bee\X\nabla_{\bee}\phi,~~~\frac{d\bee}{d\tau}=\bj\X\nabla_{\bee}\phi+\bee\X\nabla_{\bj}\phi,
\end{equation}
where
$\tau=t/t_{\rm sec}$ and $t_{\rm sec}={\sqrt{GMa}}/{\Phi_0}$ is the secular timescale. Physical solutions are
restricted to those satisfying the physical constraints $j^2=|\bj|^2=1-e^2,$ and $\bee\cdot \bj=0$.

\paragraph*{Kozai-Lidov Cycles}
When expanded to the quadrupole order only (i.e., $\ep_{\Oct}=0$), the averaged perturbing potential is axisymmetric. As a consequence, $j_z$ is conserved
and the equations of motion are invariant under rotational transformations 
around the $z$ axis. In this case, $e$, $i$, $\om$ and $i_e$ undergo periodic oscillations (Kozai-Lidov cycles), which are determined by the two constants of motion $j_z$ and 
$\phi = \phi_\Quad$ \cite[e.g.][]{Lidov62, Kozai62}. It is convenient to 
use the constant of motion 
$C_K = \frac43 \phi_\Quad - \frac12 j_z^2 + \frac16$, which is given by
\begin{equation}
C_K = e^2 - \frac52 e_z^2 = e^2 (1 - \frac52 \sin^2 i \sin^2 \omega).
\end{equation}
When $C_K < 0$, $\om$ librates around $\pi/2$ or $-\pi/2$ 
(Kozai-Lidov librations), while for $C_K > 0$, $\om$ varies monotonically with time 
taking all values from $0$ to $2\pi$ (Kozai-Lidov rotations).

For rotations, $e$ reaches minimum at $\om=0$ or $\pi$, 
implying   
\begin{equation}
\emt = C_K~~~~~(\text{only when}~C_K>0).
\end{equation}
For librating solutions, $e_{\min}$ is obtained at $\om=\pm\pi/2$. For 
any Kozai-Lidov cycle, maximum $e$ is obtained at $\om=\pm\pi/2$, 
leading to
\begin{equation}\label{eq:emax}
3e^4_{\max}+(5j_z^2-3+2C_K)e^2_{\max}-2C_K=0.
\end{equation}

To fully specify the trajectory, the azimuthal position must be specified.  We use the azimuthal angle $\Om_e$, which, as shown below changes slowly for the regime $j_z^2\ll1$ of interest here. In fact, the equation of motion for $\Om_e$ reads 
\begin{equation}\label{eq:dotOme}
\dot\Om_e = j_zf_{\Om}
\end{equation} 
where 
$
f_{\Om}=3(8-6/\sin^2i_e)/8.
$
\paragraph*{Kozai-Lidov Cycles with $j_z=0$}
In this case, Eqs. \eqref{eq:phi_Q_phi_oct}, and \eqref{eq:NLEOM} imply that the torque is given by $\dot\bj=-(15/4)e_z\bee\times\hz$, and is directed (up to sign) in the direction of $\bj$. This means that $\bj$ moves on a straight line through the origin $\bj=0$ (at which $e=1$) in the $xy$ plane periodically with
\begin{equation}\label{eq:djdt_jz0}
\dot j|_{j_z=0}=\pm \frac{15}{8} e^2\sin 2i_e.
\end{equation} 
Each cycle, when $\bj$ crosses the origin, $\Om$ jumps by $\pi$ ($\Om$ is constant otherwise).  
 
$\bee$ is confined to the plane perpendicular to the line on which $\bj$ moves and thus $\Om_e$ is constant as long as $\bee$ does not cross $\hz$. For $\bee$ to cross $\hz$, $e$ must equal $e_z$ so that $C_K<0$. Hence, for rotating cycles ($C_K>0$), $\bee$ never crosses the $\hz$ axis and $\Om_e$ is constant throughout the cycle. For librating cycles, $\bee$ crosses $\hz$ each cycle changing $\Om_e$ by $\pi$. Note that the derivative of $\Om_e$, given by  Eq. \eqref{eq:dotOme}, is zero for $j_z=0$ except for a divergence at the point $i_e=0$, where $\bee$ points in the $\hz$ direction.

\paragraph*{Equations of motion} 
We next study the evolution of the system including the contribution of the octupole part of the perturbing potential. 
Since the octupole contribution is small, we can assume that at any time, 
on short time scales (of order $t_{\rm sec}$), the solutions are nearly Kozai-Lidov cycles.
The properties of these cycles at any given time are determined by the values of $j_z$ and $C_K$ at that time. 
Moreover, given that the total potential $\phi$ is conserved,
we have to a good approximation $\phi_{\Quad}=const$, implying that 
\begin{equation}\label{eq:constQuad}
C_K+\haf j_z^2=const ,~~\soo \dot C_K=-j_z\dot j_z.
\end{equation}
Thus, the problem can be reduced to finding $j_z(t)$. 

The time derivative of $j_z$ arises from the octupole potential alone; using Eqs. \eqref{eq:phi_Q_phi_oct}, and \eqref{eq:NLEOM}, it is given by
\begin{equation}
\dot j_z=\frac{75}{64}\ep_{\Oct}\left[2 j_y j_z e_z -e_y(\inv{5}-\frac85e^2+7e_z^2-j_z^2)\right].
\end{equation}

We focus on Kozai-Lidov cycles with $j_z^2\ll1$ and study the long term behavior of the system. 
Taking the lowest order terms in $j_z$, we have
\begin{equation}\label{eq:dotjz}
\dot j_z= -\ep_{\Oct}\sin(\Om_e)f_j(e,i_e)
\end{equation}
where
$
f_j=(75/64)e\sin i_e\left[\inv{5}-e^2(\frac85-7\cos^2 i_e)\right].
$

For rotating cycles, $\Om_e$  varies slowly implying that $\dot j_z$ changes slowly. For librating cycles, $\Om_e$ changes by nearly $\pi$ each cycle, implying that $\dot j_z$ goes to approximately $-\dot j_z$ and the contribution to $j_z$ vanishes to zeroth order in $j_z$ (see example in figure \ref{fig:Lib1}). 
Below we focus on rotating cycles for which $j_z$ can monotonically change over several secular timescales.  

\begin{figure}
\includegraphics[scale=0.7]{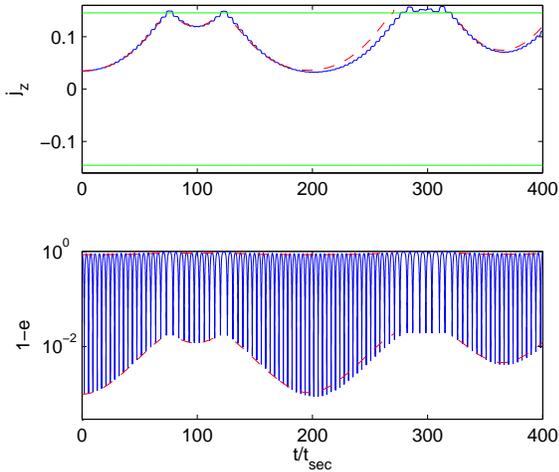}
\caption{\label{fig:t2}Results of numerical integrations for  initial conditions $\ep_{\Oct}=0.01,~\om_0=0,~\Om_0=0,i_0=88^o,e_0=0.1$. The blue solid line is the result of the direct integration of Eqs. \eqref{eq:NLEOM} while the red dashed line is the result of integrating Eqs. \eqref{eq:dOmdjzA},\eqref{eq:Avef} and \eqref{eq:constQuad} in the regions were $C_K>0$ (corresponding to $|j_z|<j_{\max}$, the region within the two green horizontal lines given by Eq.\eqref{eq:jzmax})}. 
\end{figure}

Using Eq. \eqref{eq:constQuad}, and assuming that the initial conditions $j_{z,0}, C_{K,0}$ are in the rotating zone ($C_{K,0}>0$), the condition for rotation is 
$-j_{z,\max}<j_z<j_{z,\max}$
where 
\begin{equation}\label{eq:jzmax}
j_{z,\max} = \sqrt{2C_{K,0}+j_{z,0}^2}. 
\end{equation}
If $j_z$ crosses this border, $\dot j_z$ changes sign during each cycle. For the examples of such cases we checked, after a few cycles $j_z$ moved away from this limit, back into the rotation region (see example in top panel of figure \ref{fig:t2}). We note that for some initial conditions, we found significant modulation (including flips) occurring for librating cycles on very long time scales $t\sim \ep_{\Oct}^{-2}t_{\rm sec}$, much greater than those studied here $t\sim \ep_{\Oct}^{-1}t_{\rm sec}$. 

\begin{figure}
\includegraphics[scale=0.7]{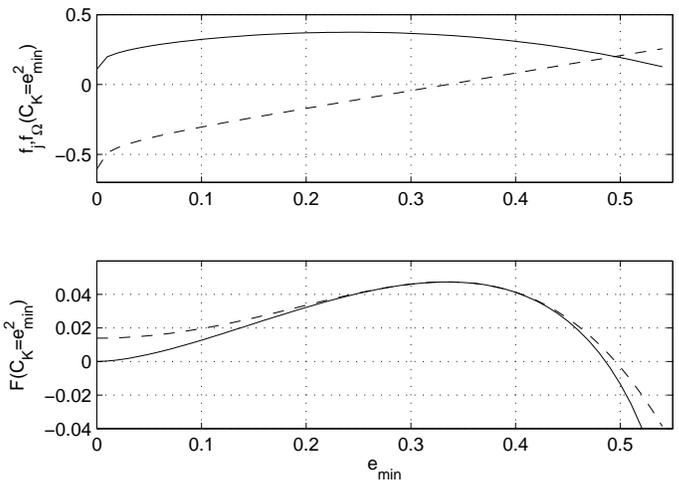}
\caption{\label{fig:fjfOm}Upper panel: $\ave{f_j}$ (solid) and $\ave{f_\Om}$ (dashed) from Eqs. \eqref{eq:Avef}. Lower panel:  $F$ from Eq. \eqref{eq:F} (solid line) and its quadratic approximation Eq. \eqref{eq:F} (dashed line)}  
\end{figure}

\paragraph*{Averaged equations}
We next average the equations of motion over the Kozai-Lidov cycle to obtain approximate  equations that describe the long term behavior of the system. 
To lowest order in $j_z,\ep_{\Oct}$, we can average Eqs. \eqref{eq:dotjz} and \eqref{eq:dotOme} over a cycle by taking the limit $j_z=0$ in which $\Om_e$ is constant and neglecting the deviation of $\dot\Om_e$ due to the octupole. The latter is important only during short episodes when $|j_z|\lesssim\ep_{\Oct}$, in which $j_z$ can change considerably during one cycle and which are not resolved in the long term approximation.
We obtain
\begin{align}\label{eq:dOmdjzA}
&\dot\Om_e=j_z\ave{f_{\Om}}\cr
&\dot j_z= -\ep_{\Oct}\ave{f_{j}}\sin(\Om_e),
\end{align}
where $f_i$ ($i=\Om, j$) are averaged over a Kozai cycle with $j_z=0$,
\begin{equation}\label{eq:average_jz0}
\ave{f_i}=\inv{\tau_{\rm Koz}}\oint_{j_z=0} dt f_i=\frac{4}{\tau_{\rm Koz}}\int_0^{\sqrt{1-e_{\min}^2}}dj\dot j^{-1}f_i,
\end{equation}
where $\tau_{\rm Koz}=\oint dt$ is Kozai-Lidov cycle period at $j_z=0$. 
To evaluate the integral, note that $e^2(1-(5/2)\cos^2i_e)=C_K$, which together with $e^2=1-j^2$ and $e_{\min}^2=C_K$ allows a straightforward integration using \eqref{eq:djdt_jz0}, yielding
\begin{align}\label{eq:Avef}
\ave{f_{\Om}}&=\frac{6E(x)-3K(x)}{4K(x)},\cr
\ave{f_{j}}&=\frac{15\pi}{128\sqrt{10}}\inv{K(x)}(4-11C_K)\sqrt{6+4C_K},\cr
x&=\frac{3-3C_K}{3+2C_K},
\end{align}
where $K(m)=\int_0^{\pi/2}(1-m\sin^2(\te))^{-1/2}d\te$ and $E(m)=\int_0^{\pi/2}(1-m\sin^2(\te))^{1/2}d\te$ are the complete elliptic functions of the first and second kind respectively.  
Note that $\ave{f_i}$ are functions of $C_K$ only since the Kozai-Lidov cycle over which the  averaging is made has $j_z=0$.
The upper panel of figure \ref{fig:fjfOm} shows plots of $\ave{f_{\Om}}$ and $\ave{f_{j}}$.

\begin{figure}
\includegraphics[scale=0.8]{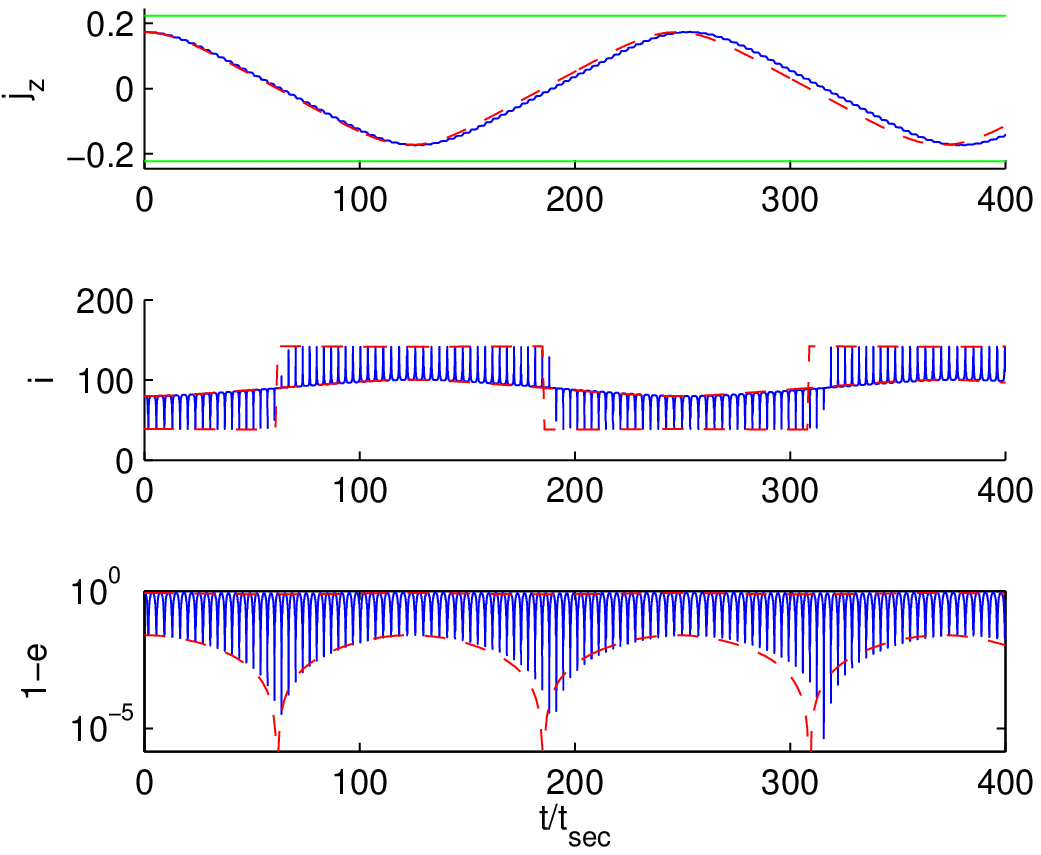}
\caption{\label{fig:t1}Results of numerical integrations for  initial conditions $\ep_{\Oct}=0.01,~\om_0=0,~\Om_0=\pi,i_0=80^o,e_0=0.1$. The blue solid lines are the result of the direct integration of Eqs. \eqref{eq:NLEOM} while the red dashed lines are the results of integrating Eqs. \eqref {eq:dOmdjzA},\eqref{eq:Avef} and \eqref{eq:constQuad} and using the pure Kozai relations to extract $e_{\min},e_{\max}$ and $i_{\min}, i_{\max}$. The two green horizontal lines in the top panel represent $\pm j_{\max}$, given by Eq.\eqref{eq:jzmax}}
\end{figure}

\begin{figure}
\includegraphics[scale=0.7]{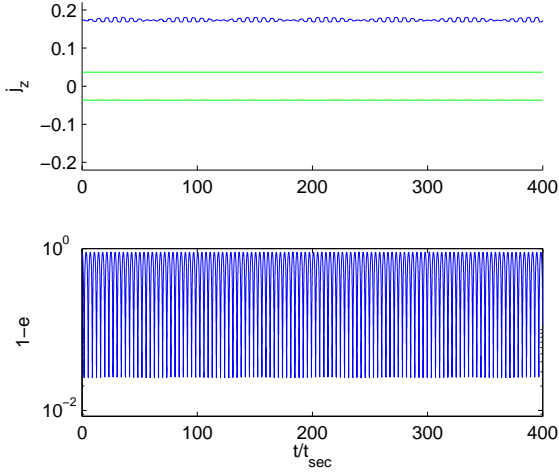}
\caption{\label{fig:Lib1}Results of numerical integrations for  initial conditions identical to the parameters in figure \ref{fig:t1} except for $\om_0=\pi/2$ replacing $\om_0=0$, for which the Kozai cycles are rotating. The blue solid line is the result of the direct integration of Eqs. \eqref{eq:NLEOM}. The two green horizontal lines represent $\pm j_{\max}$, given by Eq.\eqref{eq:jzmax}}
\end{figure}

Eqs. \eqref {eq:dOmdjzA},\eqref{eq:Avef} and \eqref{eq:constQuad} form a closed set of equations for the slowly varying $j_z,C_K$ and $\Om_e$. 
An example of a numerical integration of these equations, compared with the results of a direct integration of the secular equations \eqref{eq:NLEOM} is shown in figure \ref{fig:t1} for $\ep_{\Oct}=0.01$. As can be seen the approximate equations describe the long term evolution to a good approximation.   

These equations break down if $\abs{j_z}$ crosses the threshold Eq. \eqref{eq:jzmax}, in which case $\Om_e$ receives kicks, $\dot j_z$ changes sign, $j_z$ moves to the rotation region after a few secular time scales and the averaged equations become valid again. An example of such behavior is seen in fig \ref{fig:t2}. In this example, the effective equations Eqs. \eqref{eq:dOmdjzA},\eqref{eq:Avef} and \eqref{eq:constQuad} were integrated only in the intervals were $C_K>0$ (dashed lines).

\paragraph*{Analytic solution} 
Eqs. \eqref{eq:dOmdjzA},\eqref{eq:Avef} and \eqref{eq:constQuad} are integrable. 
These equations have a constant of motion of the form
\begin{equation}\label{eq:Constant}
C=F(C_K)-\ep_{\rm oct}\cos(\Om_e).
\end{equation}
Indeed, using \eqref {eq:dOmdjzA} and \eqref{eq:constQuad} we find, 
$
\dot C=\ep_{\rm oct}j_z\sin\Om_e(F'\ave{f_{j}}-\ave{f_{\Om}}),
$
and by setting 
\begin{align}\label{eq:F}
F(C_K)&=\int_0^{C_K}\frac{\ave{f_{\Om}}(c)}{\ave{f_{j}}(c)}dc\cr
&=32\frac{\sqrt{3}}{\pi}\int_{\frac{3-3C_K}{3+2C_K}}^{1}\frac{K(x)-2E(x)}{(41x-21)\sqrt{2x+3}}dx
\end{align}
we obtain $\dot C=0$. The numerical value of $F(e_{\min}^2)$ as a function of $e_{\min}$ is shown in the bottom panel of  figure \ref{fig:fjfOm} and tabulated in the attached file e2F.txt. 

Note that $F$ diverges at $e_{\min,\inf}=(4/11)^{0.5}$ where $\ave{f_{j}}=0$. For $e_{\min}>e_{\min,\inf}$, the integration limits in Eq. \eqref{eq:F} should be chosen differently. Here we focus on $e_{\min}<e_{\min,\inf}$. $F$ has a maximum $F_{\max}\approx 0.0475$  at $e_{\min,m}^2=(3-3x_m)/(3+2x_m)\approx0.112$ where $x_m$ is the solution to the equation $K(x_m)=2E(x_m)$ (at which $\ave{f_{\Om}}=0$). Near maximum, $F$ can be well approximated by a quadratic expression, 
\begin{equation}\label{eq:F_approx}
F(e_{\min}^2)\approx F_{\max}-2.67(e_{\min}^2-e_{\min,m}^2)^2,
\end{equation}
(shown as dashed line in the bottom panel of figure \ref{fig:fjfOm}).

This constant of motion holds most of the information about the system.

\paragraph*{Flip criterion}
We next use the constant of motion Eq. \eqref{eq:Constant} to derive a criterion of the initial conditions which is necessary in order that $j_z$ "flips" i.e.  changes sign (implying $i$ goes above $90^o$ and the orbit becomes retrograde relative to the perturber).

\begin{figure}
\includegraphics[scale=0.7]{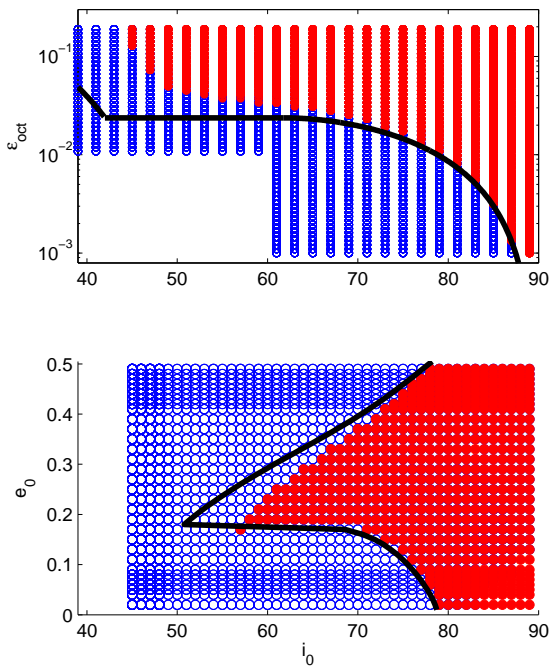}
\caption{\label{fig:ep_i_flip_scan} Upper Panel: The solid black line is the threshold, Eq. \eqref{eq:fliprequirement}, for a flip given that we start with $e_0\sim0$ (this reduces to \eqref{eq:flipe0} for $i>61.7^o$). Red circles are results of numerical integrations ($e_0=0.001$, $\om_0=0$, $\Om_0=0-2\pi$) that had a flip while blue circles are simulations that didn't.  
Lower Panel: The solid black line is the flip threshold, Eq. \eqref{eq:fliprequirement}, for $\ep_{\Oct}=0.01$. Red circles are results of numerical integrations ($\om_0=0$, $\Om_0=0-2\pi$) that had a flip while blue circles are simulations that didn't. }\end{figure}

\begin{figure}
\includegraphics[scale=0.7]{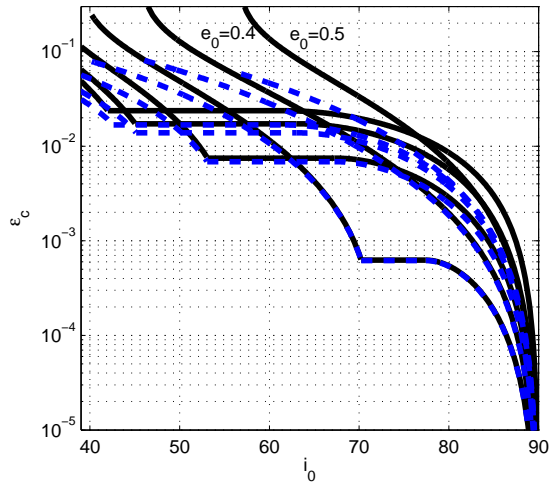}
\caption{\label{fig:ep_crit_th} 
The theoretical threshold Eq. \eqref{eq:fliprequirement}, for a flip for different initial $e_{\min}$ ($e_0$ at $\om=0$). The black solid lines show the result of Eqs. \eqref{eq:fliprequirement} and \eqref{eq:F}, for $e_{\min}=0.001,0.1,0.2,0.3,0.4,0.5$ (lines in the top left corner are ordered low $\ep_{\Oct,c},i_0$ to high $\ep_{\Oct,c},i_0$  for growing $e_{\min}$). The dashed blue curves are the corresponding results of Eq.\eqref{eq:fliprequirement}, with the approximate expression for $F$, Eq. \eqref{eq:F_approx}. }
\end{figure}

During a flip, $j_z=0$ and Eq. \eqref{eq:constQuad} implies that $C_K=C_{K,0}+0.5j_{z,0}^2$. Given the constant of motion Eq. \eqref{eq:Constant}, and that the term $\ep_{\Oct}\cos\Om_e$ can change by at most $2\ep_{\Oct}$, a required condition for a flip is that $\ep_{\Oct}>\ep_{\Oct,c}$ where 
\begin{equation}\label{eq:fliprequirement}
\ep_{\Oct,c}=\frac12\max\left(\abs{\Delta F(x)}\right)
\end{equation}
where $x$ is in the range $C_{K,0}<x<C_{K,0}+0.5j_{z,0}^2$ and $\Delta F(x)=F(x)-F(C_{K,0})$. Given that $F$ has one maximum at $C_{K,m}=e_{\min,m}^2$, the condition can be separated as follows:
If $C_{K,0}+0.5j_{z,0}^2<C_{K,m}$, or $C_{K,0}>C_{K,m}$, so that the initial and final $C_K$ are on the same side of the location of $C_{K,m}$, $\ep_{\Oct,c}=\abs{F(C_{K,0}+0.5j_{z,0}^2)-F(C_{K,0})}$. Otherwise, if the initial and final $C_K$ are on different sides of $C_{K,m}$,  the value of $\ep_{\Oct}$ is the larger of $\ep_{\Oct,c}=\abs{F(C_{K,0}+0.5j_{z,0}^2)-F(C_{K,0})}$ and $\ep_{\Oct,c}=F_{\max}-F(C_{K,0})$. The presence of $F_{\max}$ thus creates a discontinuity in the flip condition (see figures \ref{fig:ep_i_flip_scan} and \ref{fig:ep_crit_th}).
For cases where initially $e_0\ll1$ implying that $C_K\ll1$ and $j_{z,0}=\cos i_0$, and for $j_{z,0}^2<2e_{\min,m}^2$ ($i_0>61.7^o$), Eq. \eqref{eq:fliprequirement} reduces  to 
\begin{equation}\label{eq:flipe0}
\epsilon_{Oct,c}=\frac12F(\frac12\cos^2i_0).
\end{equation} 

This analytic theoretical line is shown in solid blue in the upper panel of figure \ref{fig:ep_i_flip_scan}.
For comparison, the results of numerical integrations of Eqs. \eqref{eq:NLEOM} over $10/\ep_{\Oct}$ secular times for corresponding $\ep_{\Oct},~i_0$ with $\om_0=0$, $e_0=0.001$ and $\Om_0$ scanned over $0-2\pi$ are shown in filled red (flipped) and empty blue (no flip) circles. The analytical curve describes the flip condition to better than 10\% for $i_0\gtrsim 80$ deg, better than 20\% for $i_0>70$ deg,  and to a factor less than $2$ for $i>50$ deg. The deviation at low inclinations is not surprising, given that our formalism assumes $j_z^2\ll1$. It is encouraging that the overall behaviour is captured quite well for $j_z$ up to 0.5. 

The curve representing Eq. \eqref{eq:fliprequirement} for $\ep_{\Oct,c}=0.01$ on the $e_{\min,0},i_0$ plane is shown in the lower panel of figure \ref{fig:ep_i_flip_scan} (solid black). For comparison, the results of numerical integrations for corresponding $e_{\min,0},i_0$ are shown in filled red (flipped) and empty blue (no flip) circles.   

The result of Eq. \eqref{eq:fliprequirement} for different values of $e_{\min}$ is shown in figure \ref{fig:ep_crit_th}.

\paragraph*{Maximal $e$ and General Relativity (GR) precession}
A rough estimate of the typical maximal eccentricity expected during an episode when $j_z$ crosses zero, can be obtained as follows and was confirmed in several numerical test runs with various parameters. The assumption is that at maximal $e$ we have $j\sim |j_z|$. 
Since the maximal $e$ will be obtained in some arbitrary phase in the Kozai cycle during which $j_z$ crosses zero, we expect $1-e_{\max}^2=j^2\sim j_z^2\sim (0.5\dot j_z\tau_{\rm Koz})^2\sim \ep_{\Oct}^2$ (where we assumed that $\Om_e$ is not tuned to $0$ or $\pi$ so that ($\sin^2\Om_e\sim1$). In fact we expect a roughly uniform distribution of $j_z$ or $[1-\max(e)]^{1/2}$ around this range.
General relativistic corrections cause the pericenter to rotate and can be incorporated into the equations of motion by adding a term $\phi_{\rm GR}=\ep_{\rm GR}/j$ to the total normalized averaged potential $\phi$\cite{Fabrycky07}, where $\ep_{\rm GR}=3GM^2/M_{\per}a^{-4}a_{\per}^3$. 
This effect suppresses the maximal eccentricity in the Kozai-Lidov cycles when $\ep_{\rm GR}/j\sim 1$ \cite{Fabrycky07};  using the estimate above for $e_{\max}$, GR becomes significant once  $\ep_{\rm GR}\sim \ep_{\Oct}$. If this occurs, $\bee$ can change its direction considerably during one Kozai-Lidov cycle, $\Om_e$ may change by $\sim \pi$, causing $\dot j_z$ to change sign avoiding a flip. We verified this rough criterion for suppression of flips numerically. 

\paragraph*{Kozai-migrated Hot Jupiters by distant, stellar mass perturbers}
Recently, numerical simulations by \citep{Naoz11} showed that Jupiter-mass planets, perturbed by comparable planetary or brown dwarf mass perturbers, undergo slowly modulated Kozai-Lidov cycles, and exhibit striking features, including extremely high eccentricities and the generation of retrograde orbits with respect to the perturber.  
Those authors attributed the slow modulation of the Kozai-Lidov cycles to the relative proximity and comparable mass of the perturber to the perturbed planet, and the high eccentricities to chaotic evolution; they suggested that these effects would not occur in the case of distant stellar mass perturbers.

Our analytical and numerical analysis shows that these effects have a simpler explanation: the octupole perturbation alone modulates the Kozai-Lidov cycles, and it does so non-chaotically to excite the extremely high eccentricities and the retrograde inclinations.  This effect occurs already in the test particle approximation in which the mass of the perturber is much bigger than that of the planet, and will thus be important for stellar mass perturbers.  In fact, similar evolution occurs in other cases where there are small deviations from axisymmetry \cite{Katz11}. 

Consequently the distribution of orbital parameters of Kozai-migrated hot Jupiters due to a stellar perturber may be significantly affected by the dynamics of the octupole perturbations described in this Letter. 

Consider for example, the system parameters studied numerically in \cite{Fabrycky07}, $a=5\AU$, $a_{\per}=500\AU$, $M=M_{\per}=M_{\odot}\sim 1000M_J$, where the contribution of the octupole term was neglected. The perturber eccentricity in this study was set to 0 for convenience (which implies zero octupole contribution), but in reality is expected to have a wide distribution of values. Such a system has an octupole coefficient, ${\ep_{\Oct}= aa_{\per}^{-1}e_{\per}/(1-e_{\per}^2)}\approx 0.01$, secular time scale of 
${\tsec \propto M^{1/2}M_{\per}^{-1}a^{-2}a_{\per}^3}= 1.8 \times 10^6\yr$, and GR precession coefficient of 
${\ep_{\rm GR}\propto M^2M_{\per}^{-1}a^{-4}a_{\per}^3}=4.7\times 10^{-5}$.

We performed a few test runs with these parameters (including GR precession but not tidal dissipation) and found that extremely high eccentricities $1-e_{\max}\sim 10^{-4}$ can be reached for inclinations $i>80^\circ$ (in reality, $e_{\max}$ would be limited by other physical effects).
For the runs we made, a flip was suppressed due to the GR precession, but for equally likely parameters with slightly closer perturbers, $a_{\rm per}\lesssim 300\rm AU$ the effect of GR precession becomes negligible and flips are also attainable in accordance with the criterion Eq. \eqref{eq:fliprequirement}.  

A numerical investigation of this problem is published simultaneously by a different group in \cite{Lithwick11}. 

We thank Scott Tremaine, Jihad Touma, Yoram Lithwick,  Fred Rasio, Smadar Naoz and Will Farr for useful discussions.  B.K. is supported by NASA through Einstein Postdoctoral Fellowship awarded by the Chandra X-ray Center, which is operated by the Smithsonian Astrophysical Observatory for NASA under contract NAS8-03060. Work by SD was performed under contract with the California Institute of Technology (Caltech) funded by
NASA through the Sagan Fellowship Program.

\end{document}